\documentclass[preprint,superscriptaddress,aps]{revtex4}
\usepackage{amsfonts}
\usepackage{graphics}
\usepackage{graphicx}
\newcommand{\be}{\begin{equation}}
\newcommand{\ee}{\end{equation}}
\newcommand{\en}{\end{equation}}
\newcommand{\ba}{\begin{eqnarray}}
\newcommand{\ea}{\end{eqnarray}}
\newcommand{\bea}{\begin{eqnarray}}
\newcommand{\eea}{\end{eqnarray}}
\newcommand{\pa}{\partial}

\def\As{A\!\!\!/}

\def\ks{k\!\!\!/}

\def\ps{p\!\!\!/}

\def\bs{b\!\!\!/}

\def\Ds{D\!\!\!\!/}
\def\ds{\partial\!\!\!/}

\begin{document}

\title{On the Lorentz-breaking theory with higher derivatives in spinor sector}
\author{J. R. Nascimento}
\affiliation{Departamento de F\'{\i}sica, Universidade Federal da Para\'{\i}ba\\
 Caixa Postal 5008, 58051-970, Jo\~ao Pessoa, Para\'{\i}ba, Brazil}
\email{jroberto@fisica.ufpb.br}
\author{A. Yu. Petrov}
\affiliation{Departamento de F\'{\i}sica, Universidade Federal da Para\'{\i}ba\\
 Caixa Postal 5008, 58051-970, Jo\~ao Pessoa, Para\'{\i}ba, Brazil}
\email{petrov@fisica.ufpb.br}
\author{C. Marat Reyes}
\affiliation{Departamento de Ciencias B\'{a}sicas, Universidad del B\'{\i}o B\'{\i}o,\\ 
Casilla 447, Chill\'{a}n, Chile}
\email{creyes@ubiobio.cl}
\begin{abstract}
We consider the two-point function of the gauge field in Lorentz-breaking theories with higher-derivative extension of the Dirac Lagrangian. We show that the Carroll-Field-Jackiw term naturally arises in this theory as a quantum correction being perfectly finite and thus displaying no ambiguities. Also, the finiteness of this term at low energy limit and the absence of large Lorentz violating corrections allows to avoid the fine-tuning problem. 
\end{abstract}

\maketitle

\section{Introduction}

The CPT- and/or Lorentz-breaking modifications of the field theory models are intensively studied now, especially in the context of the possible extensions of the standard model \cite{KostColl}. Up to now, the issues related to the possible extensions of the purely gauge sectors were more considered. The first known Lorentz-breaking term is the Carroll-Field-Jackiw (CFJ) term proposed and explicitly calculated in \cite{CFJ}. Further, other Lorentz-breaking terms were studied. A very large review on the possible Lorentz-breaking modifications of field theory models is presented in \cite{Kost}. These modifications were studied in numerous papers at the classical \cite{class} and quantum \cite{quant,aether} levels.

Further extension of study of the Lorentz-breaking theories occurred when the higher-derivative Lorentz-violating (HD LV) extensions began to be considered. The interest to these theories is called by the nontrivial behaviour of waves in these theories, f.e. in \cite{MP} where the first example of such an extension has been proposed, such solutions were shown to display a rotation of a plane of polarization in a vacuum. The systematic consideration of HD LV theories has been firstly presented in \cite{KostMew} where some HD LV modifications of the gauge sector of the QED were reviewed, both CPT-odd and CPT-even ones, and many issues related with propagation of waves in such theories were discussed. Further, some of such extensions were shown to emerge as quantum corrections \cite{ourHD}. The aspects of these theories related with causality, unitarity and stability have been discussed in \cite{CMR}.

As a continuation of these studies, recently the HD LV extensions in the fermionic sector of the QED began to be proposed. In the seminal paper \cite{KM}, some possible HD LV extensions of the fermionic sector of QED have been proposed, and their tree-level properties, such as dispersion relations, structure of irreducible constant coefficients, nonrelativistic, ultrarelativistic limits, some exact solutions and some estimations for the LV coefficients, have been discussed. Therefore, the natural continuation of this study would consist in treating the perturbative corrections generated by such extensions of QED, in particular, in studying of possibilities for generating of the known Lorentz-breaking terms such as, for example, the CFJ term, the aether term and other terms \cite{Kost}. 

In addition, when the Lorentz-breaking four-vector in these models has 
a time component, there are more degrees of freedom with some of them
connected to negative norm states,  so it is natural to expect that, following this way, one can construct the Lorentz-breaking extension of the Lee-Wick theory \cite{LW}.
In fact, radiative corrections arising due to higher-order operators with Lorentz violation
are far from being completely 
understood. The renormalization of such theories requires to control 
possible large Lorentz violations \cite{collins} and to define consistently 
the renormalization points, which in a Lorentz violating theory may be 
subtle \cite{renorm_LIV}, since the poles, and therefore the asymptotic states are essentially affected by the quantum corrections in the case of a non-zero Lorentz violation. The large Lorentz violations are more probably to 
originate from a theory containing higher-order operators since both
transmuted lower dimensional operators and higher-order loop corrections have to be properly rescaled with the high 
energy scale $\Lambda$ at which the Lorentz breaking takes place. This eventually leads to the possible unwanted
fine-tuning of the physical parameters.
Calculation of the quantum corrections arising due to the higher-order operators and consideration of their possible impact for the above mentioned fine-tuning is just the problem we consider in this paper using the methodology earlier developed and applied in \cite{quant}. However, in this paper we find that in our theory the CFJ term (and, evidently, higher Lorentz-violating terms) is superficially finite and suppressed, therefore, the fine-tuning problem simply does not arise.

The structure of the paper looks like follows. In the section 2, we formulate the higher-derivative Lorentz-breaking spinor QED and write down its propagator and vertices. In the section 3, we calculate explicitly the contributions to the CFJ term. In the summary, we discuss our results.

\section{Higher-derivative spinor QED}

We start with the following higher-derivative extension of the spinor QED in 4D:
\bea
\label{HDE}
S=\int d^4x\bar{\psi}\Big(i\Ds(1-\alpha\frac{\Ds^2}{m^2})+\bs\gamma_5+\xi(b\cdot D)^2-m\Big)\psi,
\eea
where $\Ds=\gamma^mD_m$, and $D_m=\pa_m-ieA_m$ is an usual gauge covariant derivative.
So, we modified the usual Lorentz-breaking extension of the QED by the higher derivatives. This action is invariant under the usual gauge symmetry. Our signature is $(+---)$.

The opened form of the Lagrangian is
\bea
L&=&\psi(i\ds(1-\alpha\frac{\Box}{m^2})+\xi(b\cdot\pa)^2+\bs\gamma_5-m)\psi+\nonumber\\
&+&e\bar{\psi}\As\psi+\bar{\psi}\Big[-\frac{\alpha}{m^2}e\As\Box-\frac{\alpha}{m^2}(i\ds+e\As)(-2ieA\cdot\pa-e^2A^2-ie(\pa\cdot A)-\frac{e}{2}\sigma^{mn}F_{mn})-\nonumber\\&-&
\xi\big(2ie(b\cdot A)(b\cdot\pa)+ieb^mb^n(\pa_mA_n)+e^2(b\cdot A)^2\big)
\Big]\psi.
\eea
Here we define $\sigma^{mn}=\frac{i}{2}[\gamma^m,\gamma^n]$.
The propagator of the spinor field in this case, in momentum space (for the Fourier transform defined as $f(x)=\int\frac{d^4k}{(2\pi)^4}e^{-ikx}\tilde{f}(k)$ so that $i\pa_mf(x)\to k_m\tilde{f}(k)$) looks like
\bea
S(k)=i[\ks(1+\alpha\frac{k^2}{m^2})-(\xi(b\cdot k)^2+m)]^{-1}.
\eea
Finding the inverse operator, we find
\bea
S(k)=i\frac{[\ks(1+\alpha\frac{k^2}{m^2})+\xi(b\cdot k)^2+m]}{k^2(1+\alpha\frac{k^2}{m^2})^2-[\xi(b\cdot k)^2+m]^2}\simeq 
i[\ks(1+\alpha\frac{k^2}{m^2})+m]\frac{1}{k^2(1+\alpha\frac{k^2}{m^2})^2-m^2}.
\eea
In principle, one can observe that this propagator has a richer set of poles in comparison with the usual Dirac propagator, so, actually, we have additional particles, and it is natural to expect that some of them possess a negative norm.

We introduce the notations:
\bea
Q(k)&=&1+\alpha\frac{k^2}{m^2}, \quad\, B(k)=\xi(b\cdot k)^2+m.
\eea
Actually, within our calculations we put $B(k)=m$, to disregard the irrelevant for us second order in $b_m$.
So, introducing the notation
\bea
\label{props}
\Delta(k)&=&[k^2(1+\alpha\frac{k^2}{m^2})^2-[\xi(b\cdot k)^2+m]^2]^{-1}\simeq[k^2Q^2(k)-m^2]^{-1}\equiv [R(k)]^{-1},
\eea
we can rewrite the propagator as
\bea
\label{sk}
S(k)=i\frac{\ks Q(k)+m}{R(k)}.
\eea

Now, let us discuss the vertices.
It is natural to consider, for the first step, the contributions of the first order in $b_m$. This allows to throw away the vertices proportional to $\xi$ (we kept the second order in $b_m$ in the quadratic part by didactic reasons, to derive the exact propagator of the spinor field), as well as the vertices involving third order in $A_m$. As a result, the reduced Lagrangian is
\bea
\label{lagred}
L&=&\psi\Big[i\ds(1-\alpha\frac{\Box}{m^2})+\xi(b\cdot\pa)^2+\bs\gamma_5-m\Big]\psi+\\
&+&e\bar{\psi}\As\psi+\bar{\psi}\Big[-\frac{\alpha}{m^2}e\As\Box-i\frac{\alpha}{m^2}\ds(-2ieA\cdot\pa-e^2A^2-ie(\pa\cdot A)-\frac{e}{2}\sigma^{mn}F_{mn})-\nonumber\\&-&e\frac{\alpha}{m^2}\As(-2ieA\cdot\pa-ie(\pa\cdot A)-\frac{e}{2}\sigma^{mn}F_{mn})
\Big]\psi.\nonumber
\eea
Then, it is clear that the coupling $\bar{\psi}\ds(\pa\cdot A)\psi$ does not contribute to the CFJ term (there is no such a contraction in the CFJ term), as well as all terms proportional to $(\pa\cdot A)$, but, perhaps, contributes to the higher-order terms. Therefore, the list of the relevant triple vertices is reduced to
\bea
L_{int3}&=&
e\bar{\psi}\As(1-\frac{\alpha}{m^2}\Box)\psi-e\frac{\alpha}{m^2}\bar{\psi}\ds(2A\cdot\pa-
\frac{i}{2}\sigma^{mn}F_{mn})
\psi,
\eea
where the derivatives act on all on the right.

Further, we denote these vertices as
\bea
V_1=e\bar{\psi}\As(1-\frac{\alpha}{m^2}\Box)\psi; \quad\, V_2=2e\frac{\alpha}{m^2}\pa_m\bar{\psi}\gamma^m(A\cdot\pa)\psi, \quad\;
V_3=-e\frac{\alpha}{2m^2}\bar{\psi}\gamma^l\sigma^{mn}F_{mn}\pa_l\psi.
\eea
Here we disregarded the derivatives of $F_{mn}$ in the $V_3$ because they do not contribute to the CFJ term. Also, we integrated by parts in $V_2$.

Also, we have one relevant quartic vertex whose form is
\bea
\label{lint4}
L_{int4}&=&e^2\frac{\alpha}{m^2}\bar{\psi}\As(2iA\cdot\pa+\frac{1}{2}\sigma^{mn}F_{mn})\psi.
\eea
These vertices and the propagator (\ref{sk}) will be used for quantum calculations.

\section{Quantum corrections to the two-point function}

Now, let us proceed with calculating the quantum corrections.
First, we consider the quartic vertices. Also, we concentrate on essentially Lorentz-breaking contributions, i.e. those ones vanishing at $b_m=0$. Since there is no $b_m$ factors in the vertices except of the usual term $\bar{\psi}\bs\gamma_5\psi$, only they arise in the propagator (\ref{props}), we see that all one-loop contributions with quartic vertices are described by the Feynman diagram depicted at Fig. 1. The insertion denoted by the $\bullet$ symbol is here and further for $\bs\gamma_5$.


\begin{figure}[!ht]
\begin{center}
\includegraphics[angle=0,scale=1.00]{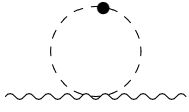}
\end{center}
\caption{Contribution with a quartic vertex}
\end{figure}

 
It is clear that in the first term of $L_{int4}$ (\ref{lint4}) there is no derivatives acting to the gauge fields. Therefore, it will yield no CFJ term. Hence we reduce our consideration to the second term of (\ref{lint4}) only.

Afterwards, the contribution from the seagull graph above turns out to be equal to
\bea
\label{sigs}
\Sigma_s(p)=\frac{e^2\alpha}{2m^2}A_l(-p)F_{mn}(p){\rm tr}\int\frac{d^4k}{(2\pi)^4}\gamma^l\sigma^{mn}\frac{1}{R^2(k)}(-\ks \bs\ks\gamma_5 Q^2(k)+\bs\gamma_5m^2).
\eea

Now, it is simple to find the trace. Remembering that $\sigma^{ab}=\frac{i}{2}[\gamma^a,\gamma^b]$, and doing the replacement $k^ak^b\to\frac{1}{4}\eta^{ab}k^2$, we get
\bea
\label{sigs1}
\Sigma_s(p)=-2\frac{ie^2\alpha}{m^2}A_m(-p)p_lA_n(p)\epsilon^{mlnb}b_b\int\frac{d^4k}{(2\pi)^4}
\frac{1}{R^2(k)}(\frac{1}{2}Q^2(k)k^2+m^2),
\eea
that is, the corresponding contribution to the effective action is given by
\bea
\label{contr0}
\Gamma_0=C_0e^2\epsilon^{ambn}b_aA_m\pa_bA_n,
\eea
where $C_0=-2\frac{\alpha}{m^2}\int\frac{d^4k}{(2\pi)^4}\frac{\frac{1}{2}Q^2(k)k^2+m^2}{R^2(k)}$ is a finite dimensionless constant depending actually only on the dimensionless parameter $\alpha$. Actually, all contributions to the CFJ term in this theory will be finite -- treating the terms with triple vertices only, the finiteness is motivated by the fact that all propagators are proportional to $\frac{1}{k^3}$, and in two vertices we have at most totally $k^4$, hence, unlike the usual Lorentz-breaking QED, the CFJ term in our case will be superficially finite.

All triple vertices also yield the CFJ-like contributions, they are depicted at Fig. 2.


\begin{figure}[!ht]
\begin{center}
\includegraphics[angle=0,scale=1.00]{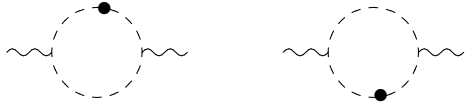}
\end{center}
\caption{Contribution with triple vertices}
\end{figure}


First, we consider the case when we have two $V_1$ vertices. 
In the standard theory, this graph has been treated in many papers, see f.e. \cite{CFJ}. In our case, the corresponding contribution is
\bea
\Gamma_1(p)&=&\frac{e^2}{2}\int\frac{d^4k}{(2\pi)^4}A_m(-p)A_n(p)Q(k)Q(k+p)
\times\nonumber\\&\times& {\rm tr}\Big[
\gamma_m S(k)\bs\gamma_5S(k)\gamma_nS(k+p)+
\gamma_m S(k)\gamma_nS(k+p)\bs\gamma_5S(k+p)
\Big].
\eea
We remind that $(1+\alpha\frac{k^2}{m^2})=Q(k)$.
So, we can write this contribution to the two-point function as
\bea
\Gamma_1(p)&=&-i\frac{e^2}{2}\int\frac{d^4k}{(2\pi)^4}A_m(-p)A_n(p)\Pi^{mn}_1(p),
\eea
where $\Pi^{mn}_1(p)$ is a self-energy tensor equal to
\bea
\Pi^{mn}_1(p)&=&{\rm tr}\int \frac{d^4k}{(2\pi)^4}Q(k)Q(k+p)\times\nonumber\\&\times&
\Big\{
\gamma^m\frac{1}{Q(k)\ks-m}\bs\gamma_5\frac{1}{Q(k)\ks-m}\gamma^n\frac{1}{Q(k+p)(\ks+\ps)-m}
+\nonumber\\&+&
\gamma^m\frac{1}{Q(k)\ks-m}\gamma^n\frac{1}{Q(k+p)(\ks+\ps)-m}\bs\gamma_5\frac{1}{Q(k+p)(\ks+\ps)-m}
\Big\}.\nonumber
\eea
Then, we expand it in power series in $p$ up to the first order, using the expression
\bea
Q(k+p)(\ks+\ps)=Q(k)k+\left(Q(k)\ps+\frac{2\alpha(k\cdot p)}{m^2}\ks\right),
\eea
and so,
\bea
\frac{1}{Q(k+p)(\ks+\ps)-m}=\frac{1}{Q(k)\ks-m}-\frac{1}{Q(k)\ks-m}(Q(k)\ps+\frac{2\alpha(k\cdot p)}{m^2}\ks)\frac{1}{Q(k)\ks-m}.
\eea
Therefore, we have, up to the first order,
\bea
\Pi^{mn}_1(p)&=&-{\rm tr}\int \frac{d^4k}{(2\pi)^4}Q^2(k)\times\nonumber\\&\times&
\Big\{
\gamma^m\frac{1}{Q(k)\ks-m}\bs\gamma_5\frac{1}{Q(k)\ks-m}\gamma^n\frac{1}{Q(k)\ks-m}(Q(k)\ps+\frac{2\alpha(k\cdot p)}{m^2}\ks)\frac{1}{Q(k)\ks-m}
+\nonumber\\&+&
\gamma^m\frac{1}{Q(k)\ks-m}\gamma^n\frac{1}{Q(k)\ks-m}(Q(k)\ps+\frac{2\alpha(k\cdot p)}{m^2}\ks)\frac{1}{Q(k)\ks-m}\bs\gamma_5\frac{1}{Q(k)\ks-m}+\nonumber\\&+&
\gamma^m\frac{1}{Q(k)\ks-m}\gamma^n\frac{1}{Q(k)\ks-m}\bs\gamma_5\frac{1}{Q(k)\ks-m}(Q(k)\ps+\frac{2\alpha(k\cdot p)}{m^2}\ks)\frac{1}{Q(k)\ks-m}
\Big\}+\nonumber\\&+&
{\rm tr}\int \frac{d^4k}{(2\pi)^4}Q(k)\frac{2\alpha(k\cdot p)}{m^2}\Big\{
\gamma^m\frac{1}{Q(k)\ks-m}\bs\gamma_5\frac{1}{Q(k)\ks-m}\gamma^n\frac{1}{Q(k)\ks-m}
+\nonumber\\&+&
\gamma^m\frac{1}{Q(k)\ks-m}\gamma^n\frac{1}{Q(k)\ks-m}\bs\gamma_5\frac{1}{Q(k)\ks-m}
\Big\},
\eea
or, as is the same,
\bea
\label{pimn1}
\Pi^{mn}_1(p)&=&-{\rm tr}\int \frac{d^4k}{(2\pi)^4}\frac{Q^2(k)}{R^4(k)}\times\nonumber\\&\times&
\Big\{
\gamma^m(Q(k)\ks+m)\bs\gamma_5(Q(k)\ks+m)\gamma^n(Q(k)\ks+m)(Q(k)\ps+\frac{2\alpha(k\cdot p)}{m^2}\ks)\times\nonumber\\&\times&
(Q(k)\ks+m)
+\nonumber\\&+&
\gamma^m(Q(k)\ks+m)\gamma^n(Q(k)\ks+m)(Q(k)\ps+\frac{2\alpha(k\cdot p)}{m^2}\ks)(Q(k)\ks+m)\bs\gamma_5\times\nonumber\\&\times&
(Q(k)\ks+m)+\nonumber\\&+&
\gamma^m(Q(k)\ks+m)\gamma^n(Q(k)\ks+m)\bs\gamma_5(Q(k)\ks+m)(Q(k)\ps+\frac{2\alpha(k\cdot p)}{m^2}\ks)\times\nonumber\\&\times&
(Q(k)\ks+m)
\Big\}+\nonumber\\&+&
{\rm tr}\int \frac{d^4k}{(2\pi)^4}\frac{2\alpha(k\cdot p)}{m^2}\frac{Q(k)}{R^3(k)}\Big\{
\gamma^m(Q(k)\ks+m)\bs\gamma_5(Q(k)\ks+m)\gamma^n(Q(k)\ks+m)
+\nonumber\\&+&
\gamma^m(Q(k)\ks+m)\gamma^n(Q(k)\ks+m)\bs\gamma_5(Q(k)\ks+m)
\Big\}.
\eea
It is clear that for $\alpha=0$ (and $Q(k)=1$), one recovers the well-known integral (see f.e. \cite{quant}):
\bea
\Pi^{mn}_1(p)&=&-{\rm tr}\int \frac{d^4k}{(2\pi)^4}\frac{1}{(k^2-m^2)^4}\times\nonumber\\&\times&
\Big\{
\gamma^m(\ks+m)\bs\gamma_5(\ks+m)\gamma^n(\ks+m)\ps(\ks+m)
+\nonumber\\&+&
\gamma^m(\ks+m)\gamma^n(\ks+m)\ps(\ks+m)\bs\gamma_5(\ks+m)+\nonumber\\&+&
\gamma^m(\ks+m)\gamma^n(\ks+m)\bs\gamma_5(\ks+m)\ps(\ks+m)
\Big\}.
\eea
So, the calculation of the trace for our integral (\ref{pimn1}) in four space-time dimensions is not very different.
After doing that, we find that the second, proportional to $\alpha$, term of (\ref{pimn1}) vanishes, and we arrive at
\bea
\label{contr1}
\Pi^{mn}_1(p)&=&-\frac{4i}{m^{10}}\int \frac{d^4k}{(2\pi)^4}\frac{Q^2(k)}{R^4(k)}\epsilon^{mnbc}b_bp_c[m^6-k^2
(\alpha k^2+m^2)^2]\times\nonumber\\&\times&
[3m^6(\alpha k^2+m^2)+\frac{3}{2}\alpha m^6k^2+\frac{\alpha}{2}k^4(\alpha k^2+m^2)^2].
\eea
It is clear that at $\alpha=0$, we reproduce the usual result \cite{quant} $\Pi^{mn}_1(p)=-4i\int\frac{d^4k}{(2\pi)^4}\frac{3m^2}{(m^2-k^2)^3}\epsilon^{mnbc}b_bp_c$. So, no singularities emerge at $\alpha=0$. Actually, it means that we can write this contribution to the effective action as
\bea
\Gamma_1&=&e^2(\frac{3}{16\pi^2}+C_1)\epsilon^{mbnc}b_bA_m\partial_cA_n,
\eea
where $C_1$ vanishes for $\alpha=0$, thus, if we treat the higher-derivative term as a regularization being removed at $\alpha=0$, the corresponding result for the Chern-Simons coefficient is $\frac{3}{16\pi^2}$, as one of the results obtained in \cite{quant}.

It remains to consider the graphs which essentially involve the vertices $V_2$ and $V_3$.
It is clear that the diagram involving two $V_3$ vertices (those ones involving $F_{mn}$) will not contribute to the CFJ term being at least of the second order in derivatives. 

Now, we consider the contraction of the vertex $V_1$ with the vertex $V_3$. The corresponding contribution is given by
\bea
\label{gamma2}
\Gamma_2(p)&=&-\frac{e^2\alpha}{2m^2}A_a(-p)F_{mn}(p)\int\frac{d^4k}{(2\pi)^4}Q(k)
\times\nonumber\\&\times&
\frac{1}{R^3(k)}{\rm tr}[\gamma^a(Q(k)\ks-m)\bs\gamma_5(Q(k)\ks-m)\ks\sigma^{mn}(Q(k)\ks-m)+\nonumber\\&+&
\gamma^a(Q(k)\ks-m)\ks\sigma^{mn}(Q(k)\ks-m)\bs\gamma_5(Q(k)\ks-m)].
\eea
The dependence of the propagators on the internal momenta is suppressed since it contributes here only to higher-derivative terms. Also, we take into account only even orders in the internal momentum $k_a$, that is, second and fourth ones, and remind that $\gamma^a\sigma^{mn}\gamma_a=0$. 

Calculating the trace, we find
\bea
\Gamma_2(p)&=&-4\frac{e^2\alpha}{m^2}A_a(-p)F_{mn}(p)\int\frac{d^4k}{(2\pi)^4}
\frac{Q(k)}{R^3(k)}k^a\epsilon^{mnbc}b_bk_c, 
\eea
or, after a simple replacement $k^ak_b\to\frac{1}{4}k^2\delta^a_b$,
\bea
\label{contr2}
\Gamma_2(p)&=&\frac{e^2\alpha}{m^2}A_a(-p)F_{mn}(p)\int\frac{d^4k}{(2\pi)^4}
\frac{Q(k)k^2}{R^3(k)}\epsilon^{mnab}b_b.
\eea
This is again the CFJ-like term. We can write it as $\Gamma_2=C_2e^2\epsilon^{mnab}F_{mn}A_ab_b$, with
the $C_2$ is also finite and dimensionless (and hence mass independent), as well as $C_0$ and $C_1$.

It remains to study the contributions essentially involving the vertex $V_2$.
There will be three such contributions -- those ones where $V_2$ is contracted with $V_1$, $V_2$, $V_3$ respectively.
The contraction of $V_2$ and $V_3$ yields
\bea
\Gamma_3(p)&=&\frac{e^2\alpha^2}{m^4}A_a(-p)F_{mn}(p)\int\frac{d^4k}{(2\pi)^4}\frac{k^ak_p}{R^3(k)}\times\nonumber\\&\times&
{\rm tr}[\gamma^p(Q(k)\ks-m)\bs\gamma_5(Q(k)\ks-m)\ks\sigma^{mn}(Q(k)\ks-m)+\nonumber\\&+&
\gamma^p(Q(k)\ks-m)\ks\sigma^{mn}(Q(k)\ks-m)\bs\gamma_5(Q(k)\ks-m)].
\eea
We see that the structure of the trace is the same as in (\ref{gamma2}), since the vertices $V_1$ and $V_2$ involve only one Dirac matrix. Therefore, we can repeat the calculation above, so, we have the following contribution from this graph to the effective action:
\bea
\label{contr3}
\Gamma_3=e^2 C_3\epsilon^{ambn}b_aA_m\pa_bA_n,
\eea
where
\bea
C_3=-\frac{\alpha^2}{2m^4}\int\frac{d^4k}{(2\pi)^4}\frac{Q(k)k^4}{R^3(k)}.
\eea
This is again the CFJ-like term, and $C_3$ is again finite and dimensionless.

The contraction of $V_1$ and $V_2$ is more complicated. It looks like
\bea
\Gamma_4(p)&=&-\frac{2e^2\alpha}{m^2}A_m(-p)A^l(p)\int\frac{d^4k}{(2\pi)^4}Q(k)k_n(k_l+p_l)
\times\nonumber\\&\times&
{\rm tr}\Big[\frac{\gamma^m(Q(k)\ks-m)\bs\gamma_5(Q(k)\ks-m)\gamma^n(Q(k+p)(\ks+\ps)-m)}{R^2(k)R(k+p)}
+\nonumber\\&+&
\frac{\gamma^m(Q(k)\ks-m)\gamma^n(Q(k+p)(\ks+\ps)-m)\bs\gamma_5(Q(k)(\ks+\ps)-m)}{R(k)R^2(k+p)}
\Big].
\eea
Following the same manner as above, we arrive at
\bea
\label{contr4}
\Gamma_4=e^2 C_4\epsilon^{ambn}b_aA_m\pa_bA_n,
\eea
where
\bea
C_4&=&8\frac{\alpha}{m^6}\int\frac{d^4k}{(2\pi)^4}\Big(\frac{5}{2}k^2(\alpha k^2+m^2)^3+3(\alpha k^2+m^2)(-3m^4\alpha k^2-3m^2\alpha^2k^4-\alpha^3k^6)+
\nonumber\\&+& \alpha m^6k^2\Big)\frac{1}{R^3(k)}.
\eea
Again, the $C_4$ is finite and dimensionless.
It is clear that it vanishes at $\alpha=0$.

Finally, we contract $V_2$ and $V_2$. In this case we have
\bea
\Gamma_5(p)&=&-\frac{2e^2\alpha^2}{m^4}A^a(-p)A^b(p)\int\frac{d^4k}{(2\pi)^4}
k_a(k_m+p_m)k_n(k_b+p_b)
\times\nonumber\\&\times&
{\rm tr}\Big[\frac{\gamma^m(\ks-m)\bs\gamma_5(\ks-m)\gamma^n(\ks+\ps-m)}{R^2(k)R(k+p)}
+\nonumber\\&+&
\frac{\gamma^m(\ks-m)\gamma^n(\ks+\ps-m)\bs\gamma_5(\ks+\ps-m)}{R(k)R^2(k+p)}
\Big].
\eea
Proceeding along the same lines, we find that it gives 0 since all relevant contributions turn out to be proportional to 
$\epsilon^{abcd}k_ak_bb_cp_d=0$ or similar vanishing expressions.

The final result for the CFJ term is given by the sum of (\ref{contr0},\ref{contr1},\ref{contr2},\ref{contr3},\ref{contr4}). It yields
\bea
\label{final}
\Gamma_{CFJ}=e^2C\epsilon^{ambn}b_aA_m\pa_bA_n,
\eea
where $C=\frac{3}{16\pi^2}+C_0+C_1+C_2+C_3+C_4$. We note that if $\alpha=0$, that is, the higher derivatives are switched off (it can be shown that for small $\alpha$ the corresponding contributions behave as $\alpha\ln\alpha$ or $\alpha^2\ln\alpha$), only the usual result for the Chern-Simons coefficient $C=\frac{3}{16\pi^2}$ survives. The constant $C$ is finite, dimensionless and mass independent. Actually, it is a function of the dimensionless parameter $\alpha$ which measures the intensity of inclusion of higher derivatives and effectively plays the role of the regularization parameter (in some sense, we can treat our study as an example of higher-derivative regularization for the calculation of the CFJ term). It is clear that, in principle, not only the CFJ-like but also aether-like similar to those ones obtained in \cite{quant} and higher-derivative contributions similar to those ones obtained in \cite{ourHD} can emerge in our theory if we take into account terms with more derivatives acting to external fields. 

\section{Summary}

We considered the higher-derivative extension of the fermionic sector of QED. We showed explicitly that within this extension, the CFJ term arises as well as in the usual Lorentz-breaking QED, however, it is explicitly finite and unambiguously determined. The value of the numerical coefficient accompanying the CFJ term essentially depends on the dimensionless $\alpha$ parameter characterizing the intensity of the Lorentz symmetry breaking and the inclusion of higher derivatives. In principle, it means that the chiral anomaly occurs also in this theory, however, to complete this study, one would need also to discuss the contribution to the effective action arising from the measure \cite{Fuji}. Also, we note that since our result for the CFJ term is explicitly finite being also finite in the absence of any regularization as occurs for the usual Lorentz-breaking QED \cite{CFJ,quant}, that is, at $\alpha=0$, we conclude that the finiteness is not a consequence of a regularization but a fundamental property of this correction, and hence, because of the fact that the finiteness stays as well when higher derivatives are switched off, the result displays no singularity at $\alpha=0$, thus we have neither fine-tuning nor large Lorentz violations unlike of some higher-derivative results obtained in \cite{MP,CMR}. It is interesting to note also that the structure of this theory is rather similar to the Horava-Lifshitz-like (HL-like) spinor QED \cite{Iengo} where one has, from one side, increasing of the degree of all denominators, from another side, arising of extra derivatives in the vertices. 

Also, since the two-point function of the spinor field evidently diverges in four dimensions in the usual QED, we can suggest that the presence of the higher-derivative extension of the free spinor field action could result in large Lorentz violation in a pure spinor sector. Indeed, the two-point function of the spinor field in our case will be superficially finite. However, the highest order in momenta in the spinor propagator is accompanied by the small parameter vanishing when the higher derivatives are switched off, therefore the result will involve this small parameter in the denominator. It may be worthwhile to find some renormalization of this theory in order to avoid fine-tuning of physical parameters. 

As a natural continuation, we can consider, first, the higher-derivative Lorentz-breaking extension of the scalar field theory, second, the higher-derivative Lorentz-breaking extension of the gravity, third, more sophisticated higher-derivative extensions of the spinor QED. We are planning to consider these problems in forthcoming papers.

{\bf Acknowledgements.} This work was partially supported by Conselho
Nacional de Desenvolvimento Cient\'{\i}fico e Tecnol\'{o}gico (CNPq). The work by A. Yu. P. has been supported by the
CNPq project No. 303438/2012-6. C. M. R. wants to thank the hospitality of the Universidade Federal da Paraiba (UFPB) and acknowledges support by Fondecyt Regular Grant No. 1140781 and by DIUBB Project No. 141709 4/R as well as the group of Fisica de Altas Energias of the Universidad del Bio-Bio.

\end{document}